\newcommand{\pt}{\ensuremath{p_T}}
\newcommand{\gev}{\ensuremath{\rm GeV}}
\newcommand{\gevc}{\ensuremath{{\rm GeV}/c}}
\newcommand{\gevcsq}{\ensuremath{{\rm GeV}/c^{2}}}
\newcommand{\met}{\ensuremath{\not\!\!\!E_T}}
\documentclass[	
  ,final            
  ]
  {aipproc}

\layoutstyle{6x9}

\begin{document}
 
\title{Search for Higgs at CDF}


\author{Shin-Shan Yu}{
  address={Fermi National Accelerator Laboratory, Batavia, IL 60510, USA},
  altaddress={On behalf of the CDF Collaboration}
}

\begin{abstract}
 We present the results on the searches for the SM and the non-SM Higgs boson 
production in $p\bar{p}$ collisions at $\sqrt{s}=1.96$ TeV with the CDF 
detector at the Fermilab Tevatron. Using data corresponding to 
300--700~$\rm pb^{-1}$, we search for the Higgs boson in various 
production and decay channels. No signal is observed, therefore, we set upper 
limits on the production cross-section times branching fraction as a function 
of the Higgs boson mass. 
\end{abstract}

\maketitle


\section{Introduction}
 In the standard model (SM), the Higgs mechanism provides masses to fundamental
 particles via electroweak symmetry breaking, which requires the 
existence of a scalar, neutral particle: the Higgs boson. However,
several open issues of SM, such as the fine tuning 
required to keep the quadratic radiative correction to the Higgs boson mass 
under control (hierarchy problem), suggest extensions of the SM. In many SM 
extensions, such as the supersymmetric model (SUSY), the 
left-right symmetric model, and the little Higgs model, there is a richer 
Higgs spectrum with additional neutral, charged, and doubly-charged Higgs 
bosons. Using data corresponding to an integrated luminosity of 
300--700~$\rm pb^{-1}$, CDF has searched for the SM Higgs and the 
doubly-charged Higgs boson in $p\bar{p}$ collisions at $\sqrt{s}=1.96$ TeV at 
the Fermilab Tevatron. All these searches include charge conjugate decays.

\section{Search for Standard Model Higgs}
Combining the results on the direct searches at LEP and the precision SM fits 
of electroweak data (excluding the low energy data), the mass of the SM Higgs 
boson is bounded in the range: 114.4--199~\gevcsq\ at 95$\%$ CL~\cite{lep}. 
At the Tevatron, the SM Higgs is mainly produced through gluon fusion with 
a cross-section of 0.1--1~pb. This direct production is 1--2 
orders of magnitude larger than the associated Higgs production, where the 
Higgs is produced with a $Z^0$, a $W^{\pm}$, or a $t\bar{t}$ pair. For 
$M_H$$>$130~\gevcsq, SM Higgs primarily decays into $W^+W^-$, 
and can be searched for cleanly in the gluon-fusion channel, using the final 
state with two leptons and missing transverse~\cite{transverse} energy (\met). 
However, for $M_H$$<$130~\gevcsq, the 
dominant decay mode is $H\rightarrow b\bar{b}$, and 
must be searched for in the associated production channel to suppress 
background from the direct production of heavy-flavor jets. The following 
subsections describe searches for the SM Higgs in three different 
production and decay channels. 


\subsection{Search for $H\rightarrow W^+W^- \rightarrow \ell^+\ell^-\nu_{\ell} 
\bar{\nu}_{\ell}$}
\begin{figure}
  \includegraphics[height=.2\textheight]{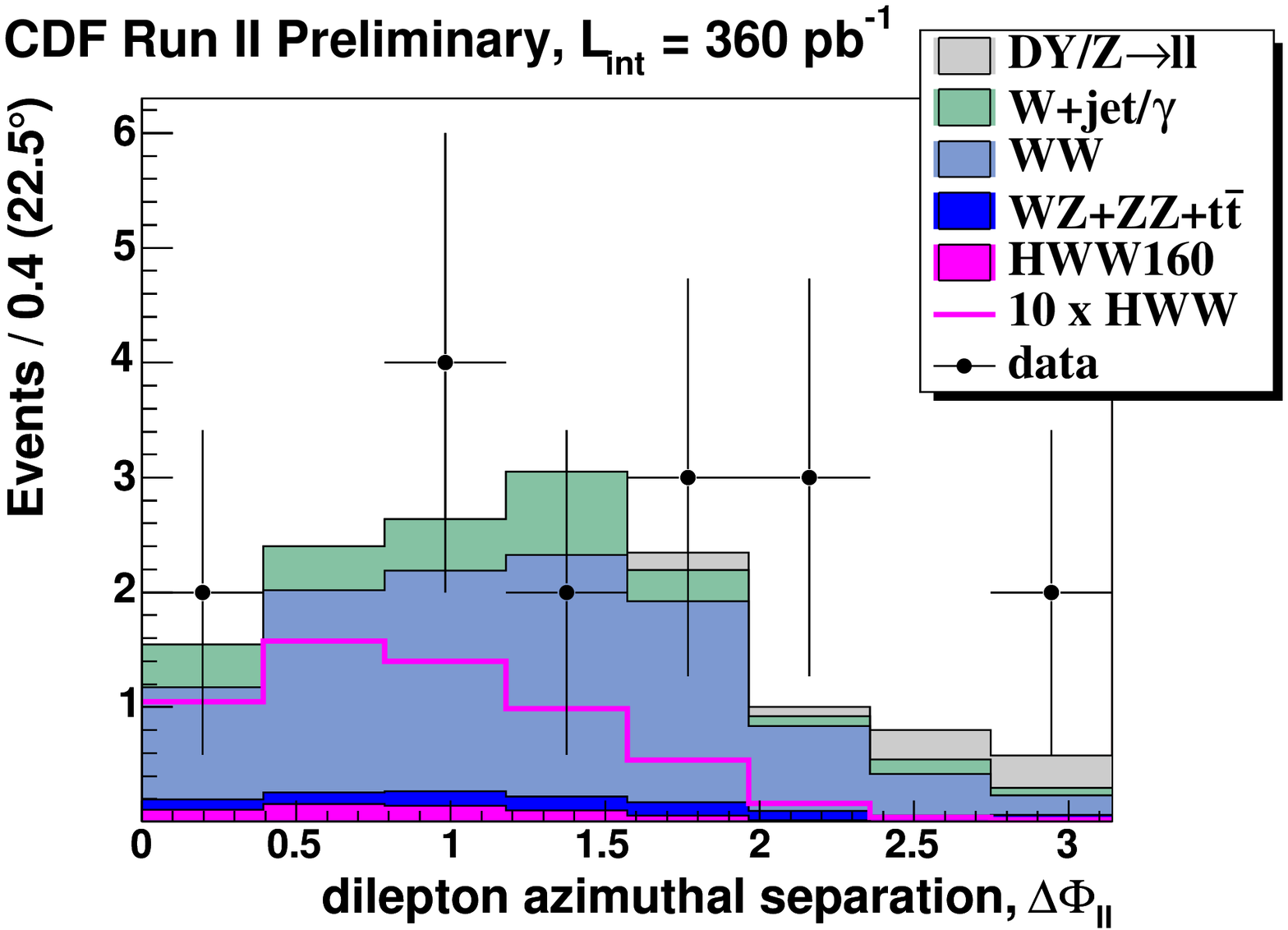}
  \includegraphics[height=.2\textheight]{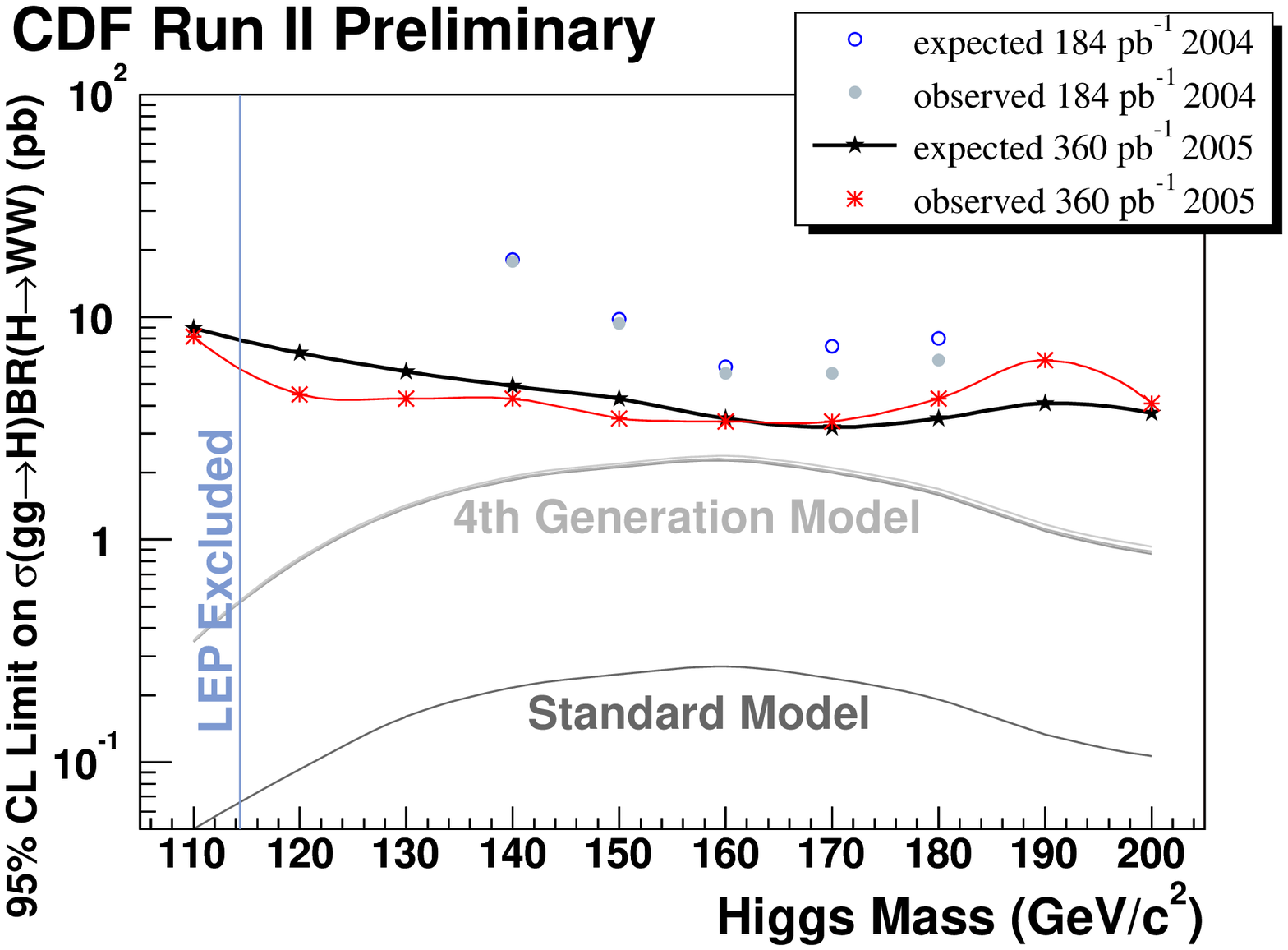}
  \caption{The $\Delta \phi_{\ell\ell}$ distributions from data and the SM 
	prediction assuming $M_H$=160~\gevcsq\ (left) and the expected and 
	observed upper limits as a function of $M_H$ (right). 
	\label{fig:hww}}
\end{figure}

This search~\cite{hww} selects events with \met$>$$\frac{M_H}{4}$~\gev\ and 
two opposite-sign isolated leptons ($e$ or $\mu$) with 
$\pt^1$$>$20 and $\pt^2$$>$10~\gevc. The decay of the spin-0 Higgs into $WW$ 
prefers a small azimuthal angle between the two leptons 
($\Delta \phi_{\ell\ell}$), a 
small dilepton invariant mass ($M_{\ell\ell}$), and large \met, and 
provides a good discriminant between signal and the SM backgrounds, especially 
the $W^+W^-$ production. In order to suppress backgrounds from $c\bar{c}$, 
$b\bar{b}$ resonances, Drell-Yan, $Z\rightarrow\tau^+\tau^-$, and mis-measured 
\met\ in addition to the SM diboson production, analysis requirements are 
applied on: $M_{\ell\ell}$ and $\pt^1$+$\pt^2$+\met\ as a function of $M_H$, 
and the angle between \met\ and the closest jet or lepton. 
Events with $\geq$1 energetic jets are also removed to reduce the 
$t\bar{t}$ background. With no evidence of a Higgs signal, a 95$\%$ CL 
upper limit on production times branching fraction is set as 
a function of $M_H$, by comparing the $\Delta \phi_{\ell\ell}$ distribution in 
data against that from the SM prediction (see Fig.~\ref{fig:hww}). 
For $M_H$=160~\gevcsq, the upper limit on 
$\sigma(gg\rightarrow H){\cal B}(H\rightarrow WW)$ is 3.2~pb.

\subsection{Search for $W^+H \rightarrow \ell^+\nu_{\ell} b\bar{b}$}
\begin{figure}
  \includegraphics[height=.25\textheight]{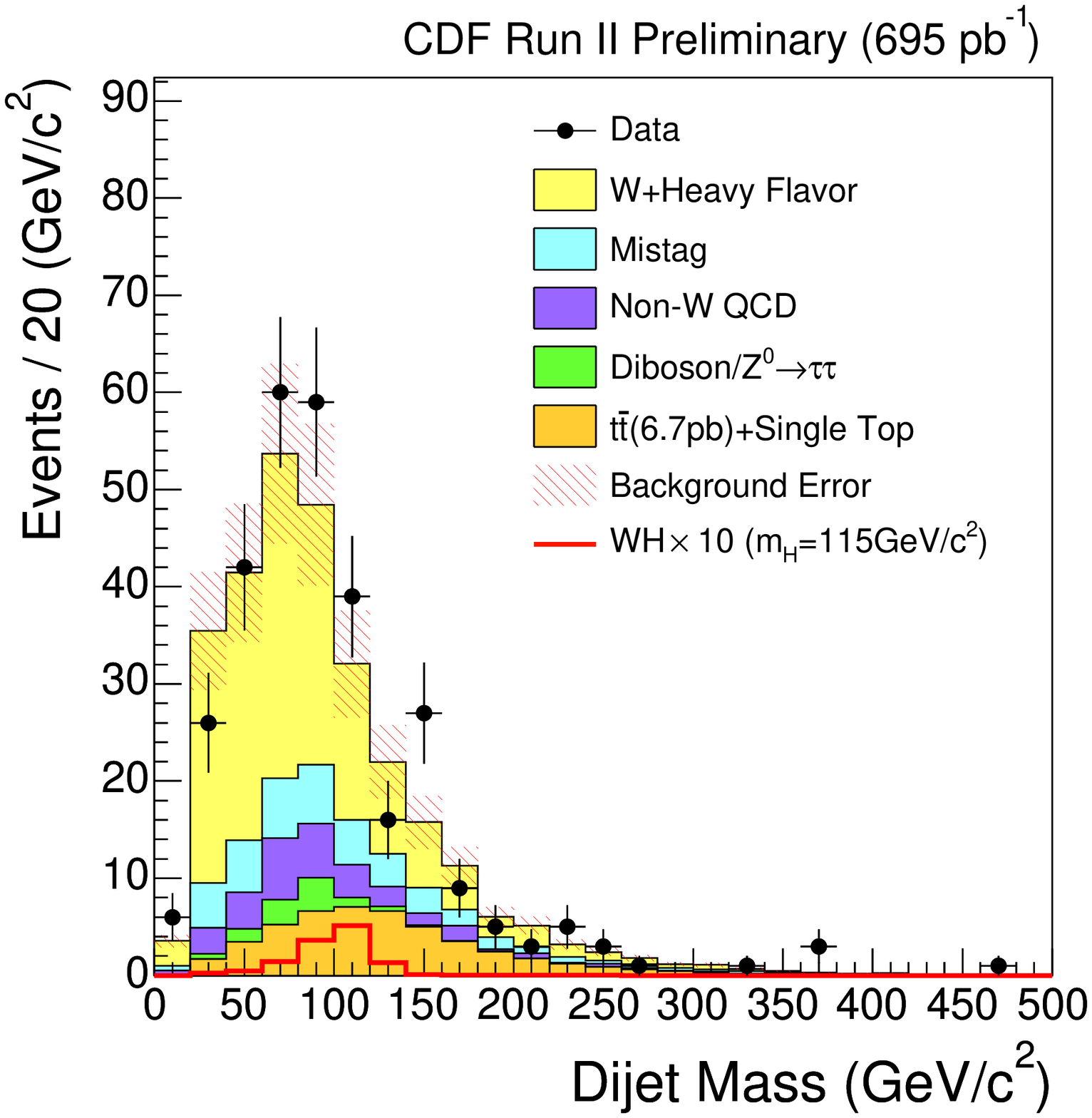}
  \includegraphics[height=.25\textheight]{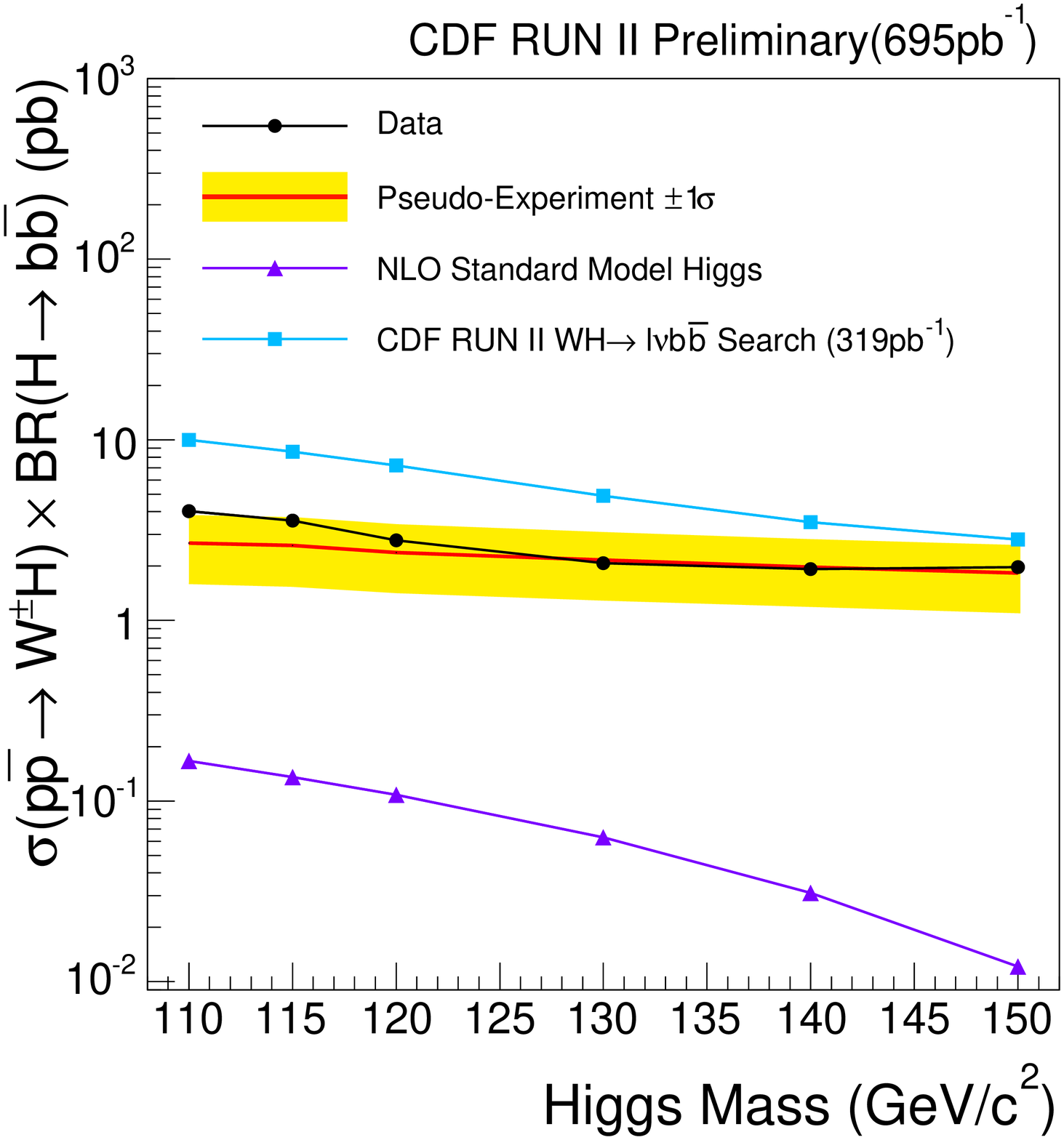}
  \caption{The dijet mass distributions from data and the SM 
	prediction for events with only one jet tagged by SecVtx and NN, 
	assuming $M_H$=115~\gevcsq\ (left) and the expected and 
	observed upper limits as a function of $M_H$ (right). 
	\label{fig:whbb}}
\end{figure}

This analysis~\cite{whbb} selects events with \met$>$20~\gev, one isolated 
lepton ($e$ or $\mu$) with \pt$>$20~\gevc, and two jets with $E_T$$>$15~\gev, 
$|\eta|$$<$2, where $\geq$1 jet must be selected by the CDF secondary vertex (SecVtx)~\cite{secvtx,Jeans:2005ew} and the neural network (NN) $b$-tagging algorithms. 
The NN has been developed to further reduce 50$\%$ of the $c$-jet and 65$\%$ of
 the light-flavor jet backgrounds while retaining 90$\%$ of the $b$-jet signal
 after SecVtx is performed. Additional requirements are applied to veto 
the Drell-Yan and $t\bar{t}$ backgrounds. The dominant background after 
$b$-tagging arises from the SM $Wb\bar{b}$ production. Events after selection 
are separated into two classes: 1. only one jet tagged by both SecVtx and NN, 
2. two jets tagged by SecVtx. By treating these two classes of events as 
independent measurements and combining their results later, the sensitivity is 
increased by $\sim$20~$\%$ with respect to that from the inclusive 
$\geq$1~$b$-tagged events. The dijet mass distribution forms a discriminant 
between signal and the SM backgrounds, and an upper limit on 
$\sigma(p\bar{p}\rightarrow WH){\cal B}(H\rightarrow b\bar{b})$ 
is set as a function of $M_H$ (see Fig.~\ref{fig:whbb}). For a Higgs boson 
mass near the LEP lower limit, $M_H$=115~\gevcsq, the upper limit is 3.6~pb.

\subsection{Search for $t\bar{t}H\rightarrow W^+W^-  b\bar{b} b\bar{b} \rightarrow \ell^+\nu_{\ell}jj b\bar{b}b\bar{b}$}
\begin{figure}
 \includegraphics[height=.2\textheight]{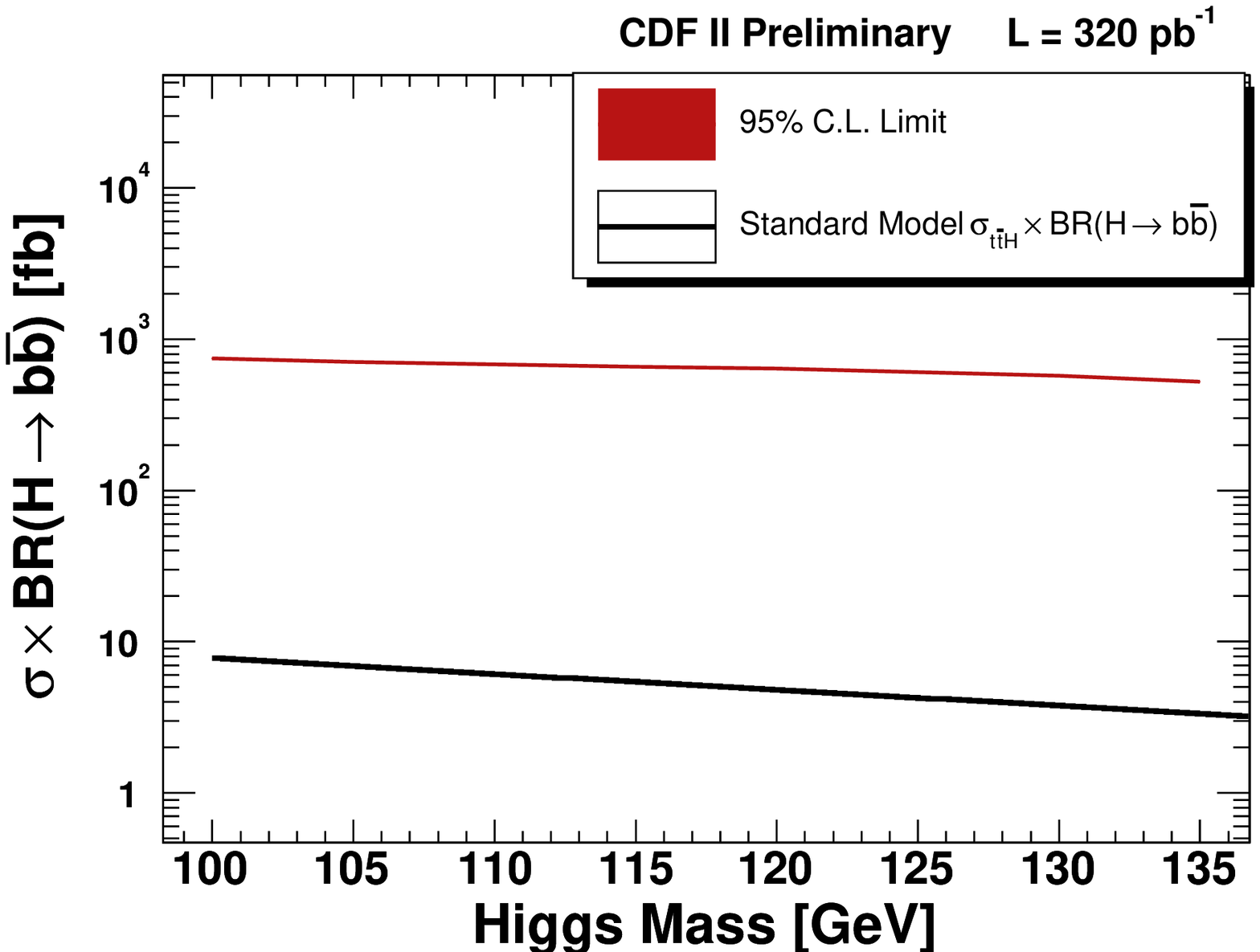}
 \includegraphics[height=.2\textheight]{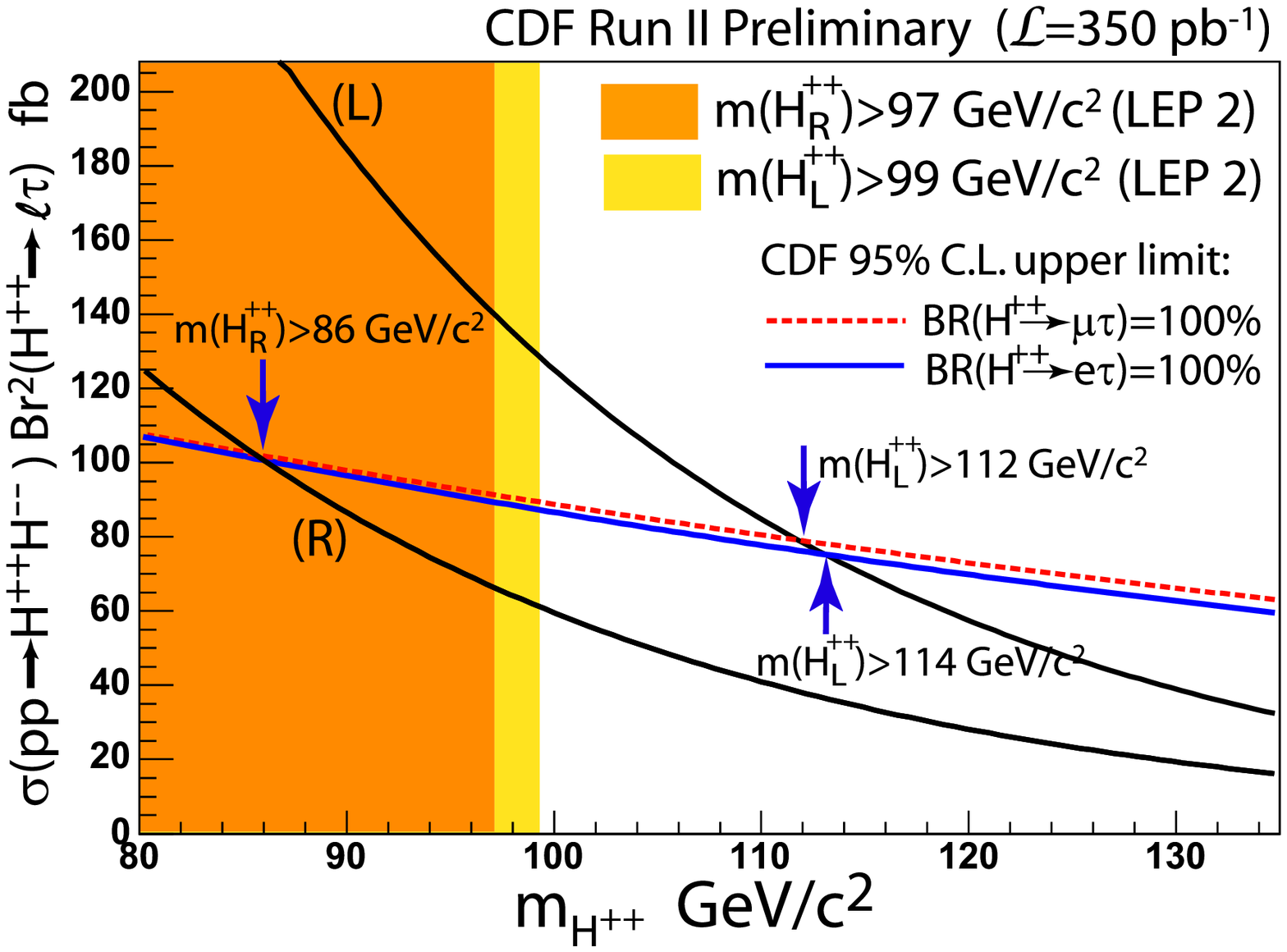}
  \caption{The observed upper limits on production times branching
	fraction as a function of $M_H$ for the $t\bar{t}H$ (left) and 
	$H^{++}$ (right) searches. The intersections of the theoretical 
	predictions and the observed limits in the right figure give lower 
	limits on the masses of left- and right-handed $H^{++}$ bosons.
	\label{fig:final}}
\end{figure}

This search~\cite{thbb} selects events with \met$>$10~\gev, one isolated 
lepton ($e$ or $\mu$) with \pt$>$20~\gevc, and $\geq$5 jets with $E_T$$>$15~\gev, $|\eta|$$<$2, where $\geq$3 jets are tagged by SecVtx. 
The leading background is the SM direct production of $t\bar{t}b\bar{b}$ and 
$t\bar{t}c\bar{c}$ events. No signal is found for the Higgs, and an upper 
limit on $\sigma(p\bar{p}\rightarrow t\bar{t}H){\cal B}(H\rightarrow b\bar{b})$
 is set as a function of $M_H$ (Fig.~\ref{fig:final}). 
For $M_H$=115~\gevcsq, the upper limit is 660~fb. Although the observed 
limit is still $\sim$2 orders of magnitude larger than the SM prediction, 
this channel is one popular channel at LHC and provides a valuable input for 
the future search.

\section{Search for non-SM Doubly-Charged Higgs}
Several extensions of SM predict the existence of the doubly-charged Higgs
 $H^{++}$. At the Tevatron, the main production mechanism is the 
pair production: $p\bar{p}\rightarrow \gamma^*/Z\rightarrow 
H^{++}H^{--}$. For $M_H$$<$160~\gevcsq, decays to $W$'s are suppressed while 
decays to leptons are theoretically unrestricted including possible 
lepton-flavor violation. This analysis~\cite{dch} searches for 
$H^{++}\rightarrow e^+\tau^+$, $\mu^+\tau^+$, which extends the previous CDF 
search for $e$ and $\mu$ final states~\cite{Acosta:2004uj}.
Analysis requires an $e$ or a $\mu$ with \pt$>$20~\gevc, 
a hadronically decaying $\tau$ with \pt$>$15~\gevc, and $\geq$ 1 
isolated track system which contains 1 or 3 tracks with $\sum \pt$$>$8~\gevc. 
Events with 3 and 4 lepton candidates are treated as independent measurements. 
In order to reduce backgrounds 
from the SM $Z$+jet, diboson, and $t\bar{t}$ production, selection criteria on 
$M_{\ell\ell}^{OS}$~\cite{os}, 
$M_{\ell\ell}^{SS}$, and $\sum\pt^{\ell}+\met$, are optimized for the 3-lepton 
and the 4-lepton events, respectively. The \met\ is also required to be at 
least 20~\gev\ for the 3-lepton events always and for the 4-lepton events only 
when $M_{\ell\ell}^{OS}$$<$120~\gevcsq. With no excess, the observed upper 
limit on $\sigma(p\bar{p}\rightarrow H^{++}H^{--}){\cal B}^2(H^{++}\rightarrow\ell\tau)$ is used to extract a lower limit on $M_{H^{++}}$, 
assuming the left-right symmetric model and exclusive decays into $e\tau$ and 
$\mu\tau$. 
The lower limits on the mass of the left-handed $H^{++}$ boson 
supersede the limit set by the LEP experiments and are found to be: 
$H_L^{++}$$>$112~\gevcsq\ for the $\mu\tau$ and 
$H_L^{++}$$>$114~\gevcsq\ for the $e\tau$ final states, respectively.

 \section{Conclusion}
CDF has searched for both the SM and non-SM Higgs bosons using 300--700~$\rm pb^{-1}$ of data. No evidence of Higgs boson production is found in the 
analyzed data, yet. As more data are being collected and more advanced analysis
 techniques are being developed, by combining the CDF results with those of D0,
 the Tevatron experiments have the potential to discover SM as well as non-SM 
Higgs.


\begin{theacknowledgments}
The author would like to thank A.~Anastassov, S.~Baroiant, S.H.~Chuang, 
T.~Junk, Y.~Kusakabe, S.~Lai, T.~Masubuchi, A.~Safonov, and A.~Taffard. 
\end{theacknowledgments}

\bibliographystyle{aipproc}   


\begin{thebibliography}{9}

\bibitem{lep}
LEP Electroweak Working Group, http://lepewwg.web.cern.ch/LEPEWWG/ (2006).

\bibitem{transverse}
The transverse plane is perpendicular to the beam line.

\bibitem{hww}
  A.~Abulencia {\it et al.}  (CDF Collaboration),
  Phys.\ Rev.\ Lett.\  {\bf 97}, 081802 (2006).

\bibitem{whbb}
Y.~Kusakabe {\it et al.}, http://www-cdf.fnal.gov/physics/exotic/r2a/20060420.lmetbj$\_$wh/ (2006).


\bibitem{secvtx}
  The SecVtx algorithm selects jets containing tracks which form a 
vertex significantly displaced from the point of $p\bar{p}$ collisions. 

\bibitem{Jeans:2005ew}
  D.~Jeans, FERMILAB-CONF-05-501-E (2005).

\bibitem{thbb}
S.~Lai {\it et al.}, http://www-cdf.fnal.gov/$\sim$slai/ttH$\_$public.html (2006).

\bibitem{dch}
S.~Baroiant {\it et al.}, http://www-cdf.fnal.gov/physics/exotic/r2a/20060406.HPlusPlus/ (2006).

\bibitem{Acosta:2004uj}
  D.~Acosta {\it et al.}  (CDF Collaboration),
  Phys.\ Rev.\ Lett.\  {\bf 93}, 221802 (2004).

\bibitem{os}
The superscript ``OS'' stands for opposite-sign 
dilepton pair, while ``SS'' stands for same-sign pair.

\end{thebibliography}



\end{document}